\documentclass[10pt,aps,pre,twocolumn,showpacs,superscriptaddress]{revtex4-1}
\usepackage{hyperref}
\usepackage{graphicx}
\begin{document}

\title{Polarization exchange of optical eigenmode pair \texorpdfstring{\\}{}
in twisted-nematic Fabry-P\'erot resonator}

\author{Vladimir A. Gunyakov}
\affiliation{Kirensky Institute of Physics, Federal Research Center -- Krasnoyarsk Scientific Center, Siberian Branch, Russian Academy of Sciences, Krasnoyarsk 660036, Russia}
\author{Ivan V. Timofeev}
\email{tiv@iph.krasn.ru}
\affiliation{Kirensky Institute of Physics, Federal Research Center -- Krasnoyarsk Scientific Center, Siberian Branch, Russian Academy of Sciences, Krasnoyarsk 660036, Russia}
\affiliation{Institute of Nanotechnology, Spectroscopy and Quantum Chemistry, Siberian Federal University, Krasnoyarsk 660041, Russia}
\author{Mikhail N. Krakhalev}
\affiliation{Kirensky Institute of Physics, Federal Research Center -- Krasnoyarsk Scientific Center, Siberian Branch, Russian Academy of Sciences, Krasnoyarsk 660036, Russia}
\affiliation{Institute of Engineering Physics and Radio Electronics, Siberian Federal University, Krasnoyarsk 660041, Russia}
\author{Victor Ya. Zyryanov}
\affiliation{Kirensky Institute of Physics, Federal Research Center -- Krasnoyarsk Scientific Center, Siberian Branch, Russian Academy of Sciences, Krasnoyarsk 660036, Russia}

\date{\today}

\begin{abstract}
The polarization exchange effect in a twisted-nematic Fabry-P\'erot resonator is experimentally confirmed in the regimes of both uniform and electric-field-deformed twisted structures. The polarization of output light in the transmission peaks is shown to be linear rather than elliptical. The polarization deflection from the nematic director grows from $0^\circ$ to $90^\circ$ angle and exchanges the longitudinal and transverse directions. Untwisting of a nematic by a voltage leads to the rotation of the polarization plane of light passing through the resonator. The polarization exchange effect allows using the investigated resonator as a spectral-selective linear polarizer with the voltage-controlled rotation of the polarization plane.
\end{abstract}


\maketitle

\section{Introduction}

In recent years, close attention of researches has been paid to the wave processes in optically anisotropic materials, including twisted liquid crystals (LCs) placed inside the Fabry-P\'erot resonator. In such structures, the ease of controlling LCs by low voltage \cite{Blinov2010bk} is combined with the high resolution of the resonator 
\cite{Patel1990,Patel1991_Silberberg}. 
They allow governing the main characteristics of the transmitted light, i.e., its intensity, phase, and polarization \cite{Timofeev2015}. 
Such structures also include more-than-one-pitch diffracting chiral layer \cite{Belyakov1992b,Abdulhalim2006,Shelton1972}, that can act as a resonator mirror \cite{Mosini1993_Tabiryan,Gevorgyan2017}.
The linear polarization is convenient to control in the Mauguin regime \cite{Mauguin1911,Yeh1999b} in a twisted nematic (TN) layer or in a twisted-nematic Fabry-P\'erot resonator (TN-FPR) \cite{Ohtera2000,Zhu_WuST2003,Zhu2002_ChoiWingKit}. This case was thoroughly theoretically investigated. As it was shown using the coupled-mode theory, the $e$- and $o$-modes corresponding to two series of the TN-FPR resonant transmission spectrum are coupled by the reflection from mirrors \cite{Ohtera2000}. First, it was analytically established that the optical-axis twisting and difference between the propagation constants of the $e$- and $o$-modes in the resonator lead to their coupling, the efficiency of which depends on the homogeneity and thickness of TN layer. It is conventional to speak about the twist modes with the elliptical polarization \cite{Yeh1999b}. Second, the elliptically polarized light reflected from a mirror changes the polarization rotation sense, which causes the coupling of twisted modes on mirrors. Therefore, one has to speak about the localized resonator modes. In this case, despite the ellipticity of the resonator modes, they remain linearly polarized at the resonator output. Numerical analysis based on an extended Jones matrix method showed that two eigenmodes at the TN-FPR boundaries are linearly polarized in the orthogonal directions \cite{Zhu_WuST2003}. Under certain conditions, their polarization directions exchange the directions longitudinal and transverse to the LC director (the polarization exchange effect). Therefore, to eliminate the mode coupling for a tunable TN-FPR intended for telecom application the incident light is preferred to be an eigenmode with rotatable linear polarization \cite{Zhu_WuST2003}. 

The aim of this study was to investigate the mode coupling effect on the polarization of TN-FPR eigenmodes with a thin TN layer, in which the Mauguin condition is broken a priori, and to confirm the possibility of controlling the linear polarization exchange effect for use of the investigated resonator as an electrically-controlled rotating linear polarizer. The measured data are compared with the results of numerical simulation by the 4$\times$4 transfer matrix method. 

\section{Experimental}

\begin{figure}
\centerline{\includegraphics[width=8.4cm]{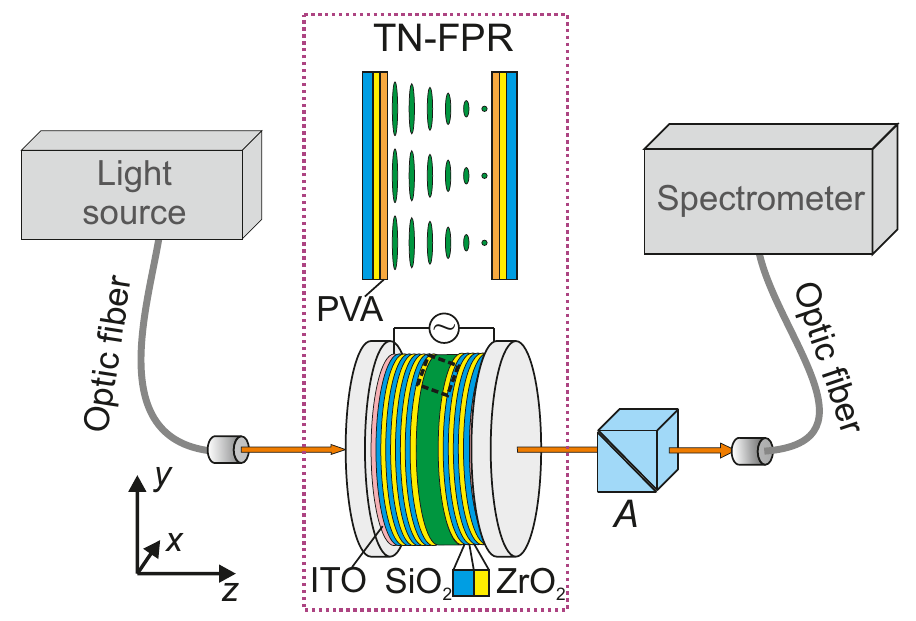}}
\caption{Schematic of the experimental setup for studying the polarization exchange effect in TN-FPR. ZrO$_{2}$/SiO$_{2}$ multilayer mirrors are formed on the substrates with transparent ITO electrodes. The resonator is filled with the twisted nematic 5CB (see inset on the top). The analyzer $A$ is a Glan prism.}
\label{fig1}
\end{figure}

The polarization states of the TN-FPR spectral transmission peaks both with and without voltage were experimentally investigated on a setup schematically illustrated in Fig.~\ref{fig1}. The Fabry-P\'erot resonator includes two identical dielectric mirrors, each consisting of six 55-nm-thick zirconium dioxide (ZrO$_{2})$ layers with a refractive index of 2.04 and five 102-nm-thick silicon dioxide (SiO$_{2}$) layers with a refractive index of 1.45 alternatively deposited onto a fused quartz substrate. The alternating ZrO$_{2}$/SiO$_{2}$ layers produce a spectral reflection band, which is the stop-band, in the spectral range of 425--625~nm \cite{Arkhipkin2008}. Thin indium tin oxide (ITO) electrodes preliminary deposited onto quartz substrates served to apply the voltage along the mirror surface normal. The resonator was assembled with a 3-$\mu $m spacers and the spacing between the mirrors was filled with a 4-n-pentyl-4$\prime $-cyanobiphenyl (5CB) LC. To form the twisted structure of LC director n, the mirrors were coated with polyvinyl alcohol (PVA) and then unidirectionally rubbed. The crossed rubbing directions ensure the uniform twisting of the nematic director n in the bulk of the LC layer by an angle of $\varphi = 90^\circ$ (Fig.~\ref{fig1}, inset). The director n at the output and input resonator mirrors is parallel to the~$x$- and~$y$-axes of the laboratory system of coordinates ($x,~y,~z)$, respectively.

\begin{figure}[hb!]
\centerline{\includegraphics[width=8.4cm]{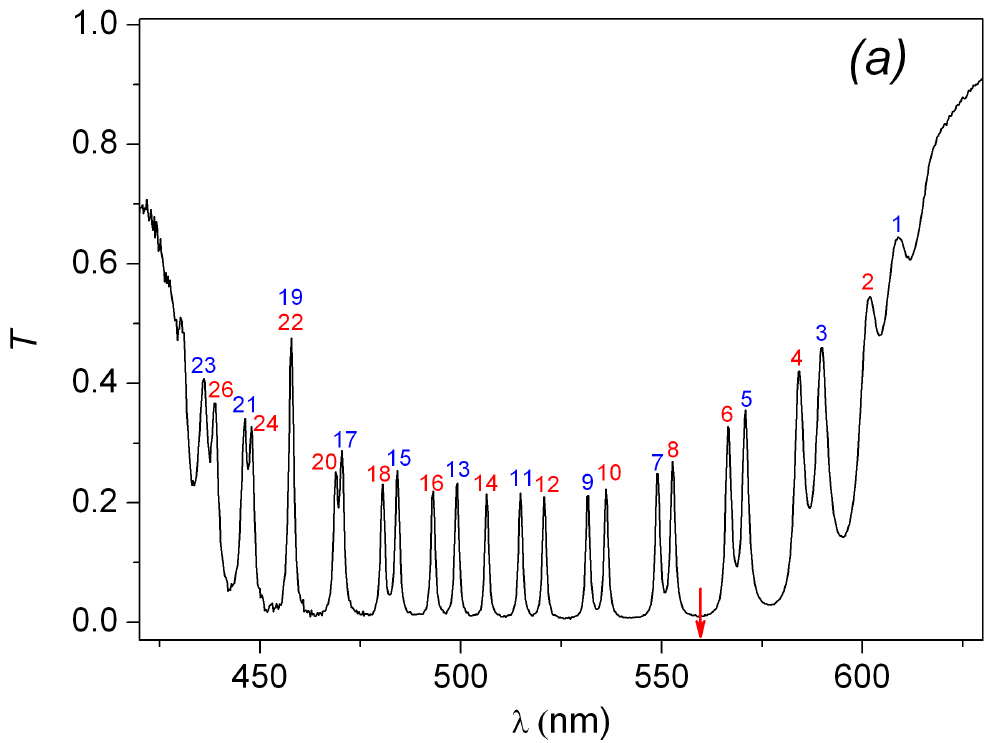}}
\centerline{\includegraphics[width=8.4cm]{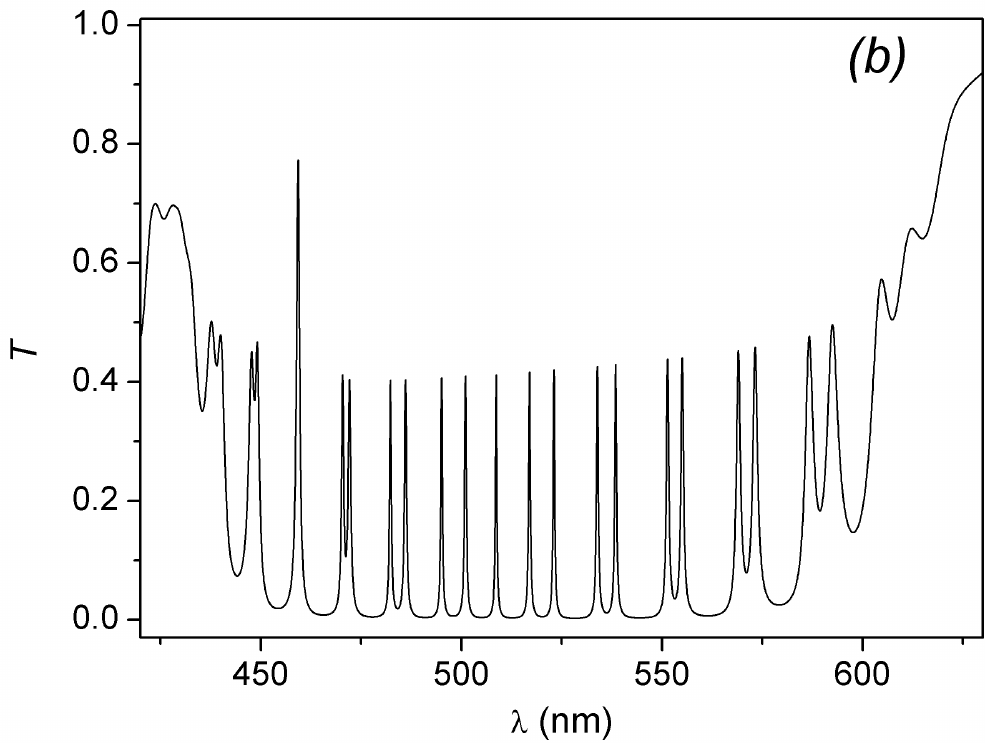}}
\caption{Spectra of the TN-FPR modes under unpolarized incident light: (a) measured and (b) simulated using the $4\times 4$ transfer matrix method with extinction 
($Im n_{LC} = 7.8 \cdot 10^{-4}$). 
The arrow shows the wavelength $\lambda_{max}=560$nm 
corresponding to the Gooch-Tarry 
maximum condition.
For convenience, the $ro$- and $re$-modes are right-to-left numbered by odd (even) numbers.}
\label{fig2}
\end{figure}

The uniform right-handed twisting of the structure was formed by resonator assembly with regard to the peculiarities of the orientant used, which specifies planar alignment of the LC director with a slight pretilt angle \cite{Kutty1983}. An ac voltage with a frequency of 1 kHz was applied to the mirrors to ensure smooth untwisting of the director n till quasi-homeotropic alignment (the twist effect \cite{Blinov2010bk}). Transmission spectra of the resonator were recorded using an HR4000 (Ocean Optics) spectrometer under unpolarized illumination at a fixed sample temperature of $t = 23.5^\circ C$. The radiation was introduced in the sample and extracted from it using light fibers. The polarization state of the resonance peaks was determined by analyzer $A$, which was a Glan prism equipped with a dial and rotating freely in the ($x,~y)$ plane.

\section{Results and Discussion} 

Figures \ref{fig2}a and \ref{fig2}b show the experimental and calculated TN-FPR mode spectrum under unpolarized incident light. The spectrum is the superposition of orthogonal transmission components, which correspond to the propagation of the resonator ordinary ($ro$) and extraordinary ($re$) modes \cite{Timofeev2015}. For convenience, the $ro$- and $re$-modes in Fig.~\ref{fig2}a are right-to-left numbered by odd (even) numbers. It can be seen that the spectrum contains two extrema: the high-intensity peak at a wavelength of $\lambda _{min}$~=~458~nm and the mode numeration inversion point at a wavelength of $\lambda _{max}$~=~560~nm (shown by the arrow). These wavelengths satisfy the Gooch-Tarry minimum and maximum conditions \cite{Gooch1975} strictly defined for Eq. (\ref{eq1}) below at the parameters of TN structure under study. The Gooch-Tarry minimum condition simulates the Mauguin regime in the LC layer for the transmitted light linearly polarized longitudinal/transverse to the director at the input. Therefore, the linearly polarized at the input $ro$- and $re$-modes coincide at $\lambda _{min}$ and degenerate. Here they add up and produce the high-intensity peak. Near $\lambda _{max}$ the mode numeration inversion is due to the mode coupling effect \cite{Zhu_WuST2003}, which leads to the phenomenon of avoided crossing \cite{Timofeev2012}. In addition, the desequencing near $\lambda _{min}$ is a trivial result of the difference between cumulative intermode spectral ranges for the $ro$- and $re$-modes, which are proportional to the refractive indices $n_{o}$ and $n_{e}$, respectively \cite{Patel1990}. The shrinkage of TN layer to a relatively small resonator thickness increases the intermode distances and repulses the neighboring Gooch-Tarry extrema to the edges of the stop-band, which makes it possible to investigate the change in the mode polarization between the extrema in a wide spectral range. 

The polarization state of every eigenmode observed in the TN-FPR transmission spectrum (Fig.~\ref{fig2}) is described by azimuth angle $\theta $ between the mode polarization direction on the output mirror and the vertical $y$-axis ($\theta = 0^\circ$). The positive angle corresponds to the deflection of polarization to the positive direction of the $x$-axis and the negative angle corresponds to the opposite side deflection. At every transmission peak the angle $\theta $ was determined by rotating the analyzer until the maximal transmission in this resonance. Upon rotation of the analyzer by $90^\circ$ from the determined angle $\theta $, we observed the total extinction of the peak, which provides evidence of the almost linear polarization of the radiation passed through the resonator.

The LC director field determines the local permittivity tensor $\varepsilon _{ij}(z)$ at all points of the medium. The extraordinary dielectric permittivity axis is collinear to the LC local director. As it was geometrically demonstrated in \cite{Timofeev2015}, the angle $\xi $ between the linear polarization on the mirror and the LC director at the input mirror is determined by the Napier's rule:
\begin{equation}
\tan 2\xi =-\sin \Theta \tan \upsilon,
\end{equation}
where $\upsilon =\sqrt {\delta ^2+\varphi ^2}$ is the twisted anisotropy phase to define Gooch-Tarry minimum at $\upsilon =\pi N$ (full-wavelength phase plate) and maximum at $\upsilon =\pi /2+\pi N$ (half-wavelength phase plate), $\Theta$ is the ellipticity factor, $\tan \Theta =\varphi /\delta$, $\varphi$ is the LC director twisting angle, $\delta = \Delta n L \omega /2 c$ is the anisotropy phase (angle), $L$ is the LC layer width, $\Delta n = n_{e} - n_{o}$ is the difference between the extraordinary and ordinary refractive indices, and $\omega $ and $c$ are the frequency and velocity of light in free space. The angle $\xi $ is defined up to a certain additive constant of $180^\circ$. For any $ro$-mode the value of angle $\xi$ is determined by equation:
\begin{equation}
\label{eq1}
\xi = \frac{\pi}{2} - \theta = \frac{1}{2}\tan ^{-1}\left[ {-\frac{\varphi }{\upsilon }\tan \upsilon } \right].
\end{equation}
Also the analytical solution of Eq. (\ref{eq1}) takes into account the implicit LC frequency dispersion. 
The $re$-mode angle differs by $90^\circ$. Expression (\ref{eq1}) is much simpler than the expressions obtained previously in \cite{Zhu_WuST2003}. Rewriting Eq. (18) from \cite{Zhu_WuST2003} in the accepted designations, we arrive at the equivalent formula
\begin{equation}
\xi =\tan ^{-1}\left[ {\frac{\cos \upsilon \pm \sqrt {1-\frac{\delta ^2\sin ^2\upsilon }{\upsilon ^2}} }{\varphi \frac{\sin \upsilon }{\upsilon }}} \right].
\end{equation}
It should be noted that in the investigated system, the angles $\xi $ (Eq.~(\ref{eq1})) and $\theta $ are complementary, i.e., their sum equals $90^\circ$. 

The TN-FPR spectrum and mode polarization were numerically simulated using the approach described in detail in \cite{Timofeev2012}. The nematic orientation structure inside the resonator with voltage was calculated by minimizing the free energy of the director field \cite{Timofeev2012,Deuling1972}. Then, using the 4$\times$4 transfer matrix method \cite{Berreman1972}, the transmission and polarization of light in the investigated multilayer structure were simulated with regard to the optical extinction in the multilayer structure and material dispersion. 

\begin{figure}
\centerline{\includegraphics[width=8.4cm]{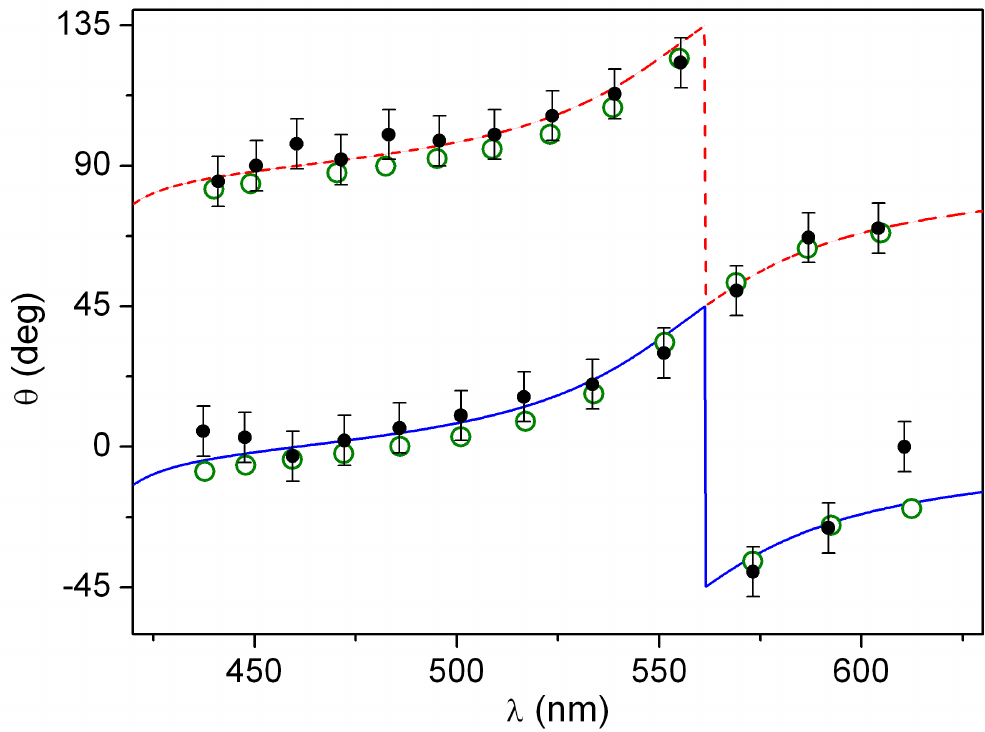}}
\caption{Polarization deflection angles $\theta $ for 26 TN-FPR transmission peaks presented in Fig.~\ref{fig2}. Closed circles show experimental values, open circles show calculated values obtained by the direct numerical simulation. Dashed and solid lines are plotted functions of the angle $\theta$ obtained from equation (\ref{eq1}); the dashed line corresponds to the resonator extraordinary $re$-modes and the solid line corresponds to the resonator ordinary $ro$-modes. }
\label{fig3}
\end{figure}

\begin{figure}
\centerline{\includegraphics[width=8.4cm]{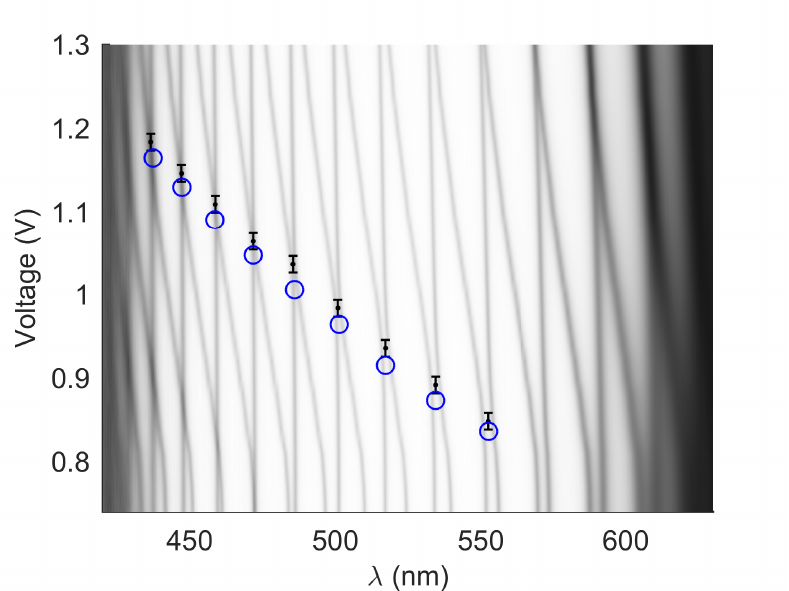}}
\caption{Experimental dependence of TN-FPR transmission on the applied voltage. The maximal and minimal transmissions are shown with black and white, respectively. The polarization exchange effect voltages at which the modes are polarized at the angles of $\pm 45^\circ$ are shown for 9 peak pairs. Closed circles show experimental values obtained by the intensity equalization technique and open circles show the calculated values.  }
\label{fig4}
\end{figure}

In the calculations using Eq. (\ref{eq1}) (Figs. \ref{fig2}b, \ref{fig3}, and \ref{fig4}) and direct numerical simulation, the following parameters were used. The mirror was a stack of the SiO$_{2}$ substrate, ITO electrode, Al$_{2}$O$_{3}$, ZrO$_{2}$, (SiO$_{2}$, ZrO$_{2})^{5}$, and PVA layers. The thicknesses and refractive indices of the amorphous layers of the dielectric mirrors were 815~nm and 1.45 for SiO$_{2}$, 66~nm and 2.04 for ZrO$_{2}$, 10~nm and 1.63 for Al$_{2}$O$_{3}$, 1.515 and 205~nm for the PVA layer. The values for the ITO layer were 117~nm and 1.88858+0.022i with account of the extinction; the substrate refractive index was 1.45 and the 5CB extraordinary and ordinary refractive indices were$ n_{II}$~=~1.737 and $n_{\bot }$~=~1.549, respectively (the wavelength is $\lambda $~=~589~nm and the temperature is $t = 24^\circ$~C); the imaginary part of the refractive index was 7.8$\cdot $10$^{-4}$i. The nematic layer thickness was 4.13~$\mu $m and the twist angle was $90^\circ$. Taking into account the dispersion, we used the following data and reference from \cite{refractiveindex_info}:
$n_{\vert \vert }(\mbox{5CB}) = 1.6808 + 0.0081\lambda ^{-2} + 0.0024\lambda ^{-4} + 0.00039i$;
$n_{\bot } (\mbox{5CB}) = 1.5139 + 0.0052\lambda ^{-2} + 0.0008\lambda ^{-4} + 0.00039i$ \cite{Li2005_5CB};
$n^{2}(\mbox{SiO}_{2}) = 1 + 0.6961663\lambda ^{2}/(\lambda ^{2} -0.06840432) + 0.4079426\lambda ^{2}/(\lambda ^{2} - 0.11624142) + 0.8974794\lambda ^{2}/(\lambda ^{2}-9.8961612)$ \cite{Malitson1965_SiO2};
$n(\mbox{ZrO}_{2}) = 1.992 + 0.02\lambda ^{-2}$ \cite{Wood1982_ZrO2};
$n(\mbox{Al}_{2}\mbox{O}_{3}) = 1.62 + 2.6 10^{-3}\lambda ^{-2} + 2.0 10^{-6}\lambda ^{-4}$ \cite{Weber2003bk};
$n(\mbox{ITO})  = 2.3479 + 0.0735i -- \lambda (0.779+0.1i)$ \cite{Konig2014_ITO};

The LC deformation under applied voltage was calculated at the following 5CB parameters ($t = 24^\circ$~C): elasticity moduli of $k_{splay}$~=~6.3~pN, $k_{twist}$~=~4.0~pN, and $k_{bend}$~=~8.9~pN and permittivities of $\varepsilon _{\vert \vert } = 18$ and $\varepsilon _{\bot } = 6$. Subscripts $\vert \vert $ and $\bot $ show the directions of the permittivity measurements (parallel and perpendicular to the director); Not only the LC, but also the mirrors were under voltage, which was taken into account as an additional voltage drop with 5{\%} of the total voltage drop before the Fredericks transition and up to 15{\%} at the homeotropic LC orientation. 

Figure \ref{fig2}b shows the calculated TN-FPR transmission spectrum at zero voltage. 
It can be seen that the experimental and calculated spectral positions of the resonator modes are in good agreement over the entire stop-band. 

Figure~\ref{fig3} shows the analyzer deflection angles for every resonant peak of the TN-FPR spectrum, which were obtained both experimentally and by direct numerical simulation. Using Eq.~(\ref{eq1}), we presented $\theta $ ($\lambda )$ in the form of a functional dependence, which emphasizes the specific features of the mode polarization. At the Gooch-Tarry minimum $\lambda _{min}$, the eigenmodes are polarized along the $x$- and~$y$-axes. Upon approaching the Gooch-Tarry maximum $\lambda _{max}$, the deflections of the linear polarizations from the axes increase up to the polarization exchange, i.e., $\theta \to +45^\circ$ for the $ro$-modes and $\theta \to +135^\circ$ for the $re$-modes. At the critical point $\lambda _{max}$, the polarization directions of both $ro$- and $re$-modes change by $90^\circ$. Then, when moving to the long-wavelength region, the polarization directions monotonically return to the corresponding axes. This is the general scenario of the eigenmode polarization state variation near the Gooch-Tarry maximum in the spectrum. 

On the other hand, changing the parameters of TN layer, e.g., decreasing the $\Delta n$ value by voltage, one can smoothly shift the Gooch-Tarry maximum to the short-wavelength spectral region and, thus, observe the change in the polarization state and induce the polarization exchange effect at any wavelength $\lambda $~$<$~$\lambda _{max}$ corresponding to the transmission peak. Figure~\ref{fig4} shows the experimental dependence of light transmission for the investigated TN-FPR on the wavelength and applied voltage. The dependence was obtained under unpolarized illumination of the sample without analyzer. The voltage was increased starting from the Fredericks threshold voltage $U_{c}$~=~0.76~V with a step of 0.02~V.

\begin{figure}
\centerline{\includegraphics[width=8.4cm]{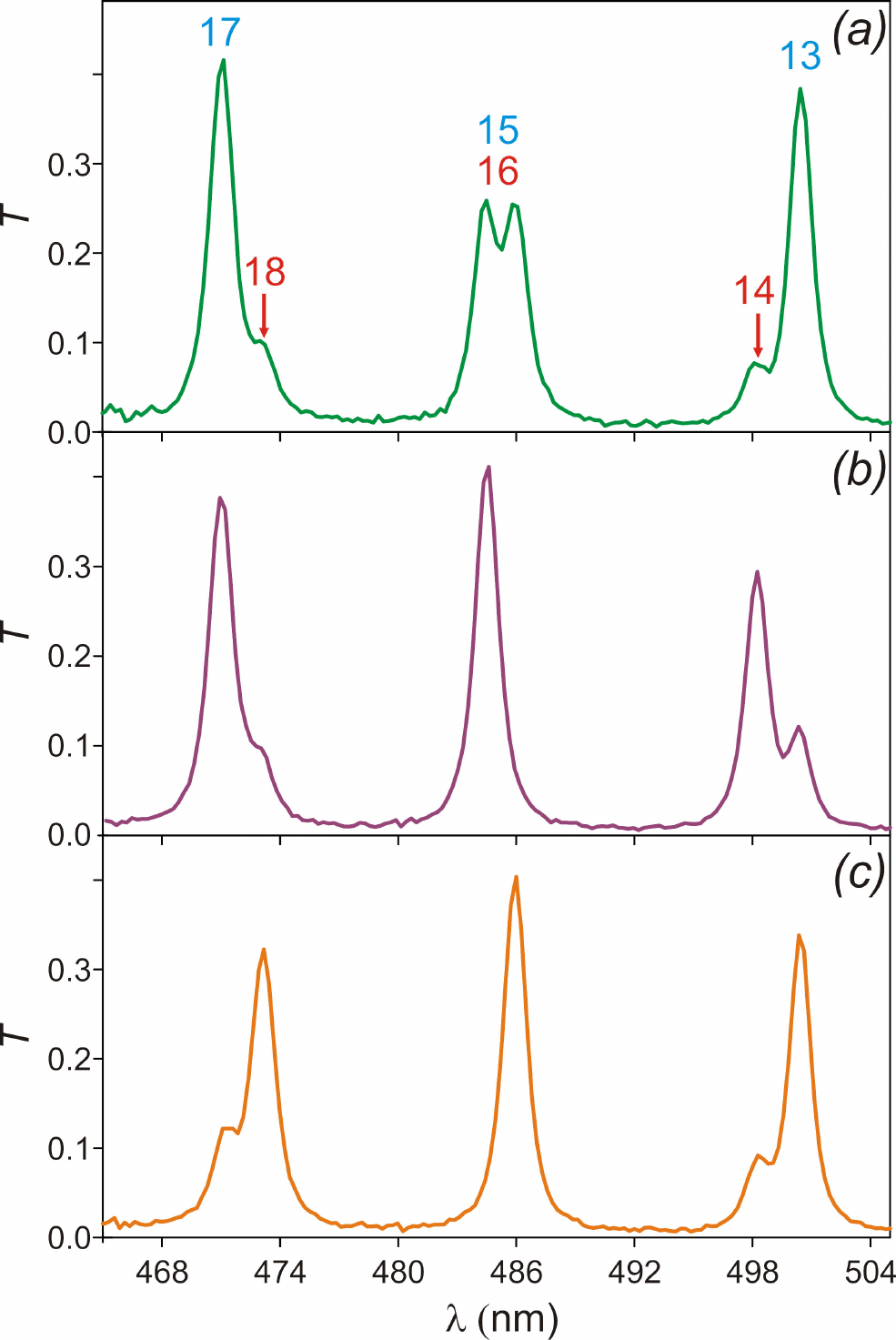}} 
\caption{Equalization of the transmission peak intensities by voltage: avoided crossing phenomenon in the vicinity of the $ro$-mode (484.5~nm) induced by a voltage of $U$~=~1.0~V without polarizer $P$ at the analyzer $A$ orientations of (a) $0^\circ$, (b) $+45^\circ$, and (c) $-45^\circ$.  
Both vertical and horizontal axes in (a-c) are at the same exact scale.}
\label{fig5}
\end{figure}

In the voltage range of 0.76-1.30~V every $ro$-mode (vertical line) undergoes at least one avoided crossing with some $re$-mode shifting to the short-wavelength region (oblique line). As an example, Fig.~\ref{fig5} shows the enlarged avoided crossing, which results from coupling of the $ro$-mode with number 15 ($\lambda $~=~484.5~nm in Fig.~\ref{fig2}) with the nearest $re$-mode with number 16 ($\lambda $~=~493.1~nm) for the three positions of analyzer $A$: along the $y$-axis and at angles of $\theta = \pm 45^\circ$. Equalization of the transmission peak intensities by a voltage at $A\parallel y$ (Fig.~\ref{fig5}a) is indicative of the polarization exchange of eigenmodes (Figs.~\ref{fig5}b and 5c). Analogously, we obtained the voltages corresponding to the polarization exchange effect for 9 pairs of peaks (closed circles in Fig.~\ref{fig4}). In addition, the calculated values of the corresponding voltages are shown (open circles in Fig.~\ref{fig4}). 

\section{Conclusions}

The polarization states of the localized resonator $ro$- and $re$-modes in the TN-FPR transmission spectrum with a priori broken Mauguin condition were studied experimentally and theoretically. 
Cavity mirrors do recover the output polarization to be always linear rather than elliptical. The linear polarization appears like in untwisted regime, but dialectically not always longitudinal to the director and still elliptical inside the TN bulk. The critical behavior of the polarization of eigenmodes with different numbers for a uniformly twisted nematic near the spectral point corresponding to the Gooch-Tarry maximum was demonstrated. The electric-field-induced deformation of twisted nematic keeps symmetric relative to the mirrors and ensures the stable polarization exchange effect \cite{FernandezPousa_Moreno2000}. This deformation leads to the shift of every $re$-mode toward the neighboring short-wavelength $ro$-mode. When it approaches the spectral position of the $ro$-mode, the Gooch-Tarry maximum occurs. The polarization direction of the $re$- and $ro$- modes monotonically changes and tends to align at angles of $\pm 45^\circ$ to nematic director. In this case, the modes remain orthogonally polarized. As the voltage increases, a sequence of avoided crossing phenomena is observed due to the strong mode coupling. The obtained experimental results were analyzed both analytically and by the numerical simulation of light transmission through the investigated multilayer structure using the 4$\times$4 transfer matrix method. The investigated TN-FPR structure can be used to create an electrically-controlled selective rotating linear polarizer. The results of this study can be generalized substituting the LC layer by smooth three-dimentional gratings \cite{Popov_Jakli2017,Nys_Neyts2015,Wang_Chigrinov2017} as well as by any structurally helical materials \cite{Faryad_Lakhtakia2014rv}.
\begin{acknowledgments}
This work was partially supported by the Siberian Branch of the Russian Academy of Sciences under Complex Program II.2P (projects Nos. 0356-2015-0410 and 0356-2015-0411). I.V.T. acknowledges the support by RFBR 17-42-240464. 
\end{acknowledgments}

\bibliography{library}
\end{document}